\documentclass[a4paper, 11pt]{article}
\usepackage{color}   
\usepackage[linktoc=all,hidelinks]{hyperref}
\hypersetup{linktocpage}
\usepackage{pdflscape}
\usepackage[textwidth=13.5cm,textheight=21.3cm]{geometry}
\usepackage{amsmath,amssymb}
\usepackage{latexsym}
\usepackage{multicol}
\usepackage{graphicx}
\usepackage{caption}
\usepackage{cancel}
\usepackage{bbm}
\usepackage{bm}
\usepackage{cite}
\numberwithin{equation}{section}
\pdfoutput=1
\usepackage[T1]{fontenc}
\usepackage[latin9]{inputenc}
\usepackage[active]{srcltx}
\usepackage{esint}
\usepackage{cancel}
\usepackage{color}
\usepackage{ulem}
\usepackage{mathtools}
\usepackage{caption}
\allowdisplaybreaks[1]

\usepackage[symbol]{footmisc}

\renewcommand{\thefootnote}{\fnsymbol{footnote}}

\newcommand{\del}{\partial}
\newcommand{\be}{\begin{equation}}
\newcommand{\ee}{\end{equation}}
\newcommand{\bea}{\begin{eqnarray}}
\newcommand{\eea}{\end{eqnarray}}

\newcommand{\nn}{\nonumber}

\newcommand{\ie}{{\it i.e.}}
\newcommand{\eg}{{\it e.g.}}
\newcommand{\ndt}{\noindent}

\usepackage{tikz}
\usetikzlibrary{trees}
\usetikzlibrary{calc}
\usetikzlibrary {arrows.meta} 
\usepackage{lipsum}
\usepackage{varwidth}
\usepackage{pifont}
\usepackage{cancel}

\numberwithin{equation}{section}
\begin{document}

		\begin{flushright}    
			{\small $\,$}
		\end{flushright}
		\centerline{\Large{\bf{On the Carrollian Nature of the Light Front}}}
		\vskip 0.5cm

		\centerline{Sucheta Majumdar\footnote{\textit{Electronic address: sucheta.majumdar@cpt.univ-mrs.fr}}}
		\vskip 0.3cm           
		\centerline{\it{Aix-Marseille Univ, Universit\'e de Toulon, CNRS,}}
			\centerline{\it{Centre de Physique Th\'eorique, 13288 Marseille, France}}
\vskip 1cm
		\centerline{\large{\bf {Abstract}}}
		\vskip 0.5cm
		\begin{center}
		\begin{varwidth}{32em}
		
\ndt  Motivated by recent advances in non-Lorentzian physics, we revisit the light-cone formulation of quantum field theories. We discuss some interesting subalgebras within the light-cone Poincar\'e algebra, with a key emphasis on the Carroll, Bargmann, and Galilean kinds. We show that theories on the light front possess a Hamiltonian of the magnetic Carroll type, thereby proposing a straightforward method for deriving magnetic Carroll Hamiltonian actions from Lorentzian field theories.
\end{varwidth}
\end{center}
\vskip 0.5cm


\renewcommand*{\thefootnote}{\arabic{footnote}}


\section{Introduction}

This article is dedicated to the memory of Lars Brink, an eminent scientist and an exceptional mentor whose scientific curiosity and wisdom profoundly influenced my perspective on fundamental physics. His valuable guidance and words of encouragement will be deeply missed, and I am forever indebted to him.
\par 
From string theory and supersymmetry to light-cone gravity and higher spins, the numerous significant contributions Lars made to light-cone physics are a testament to his keen interest in the subject. As a tribute to his deep appreciation for life on the light front, this article explores some aspects of the light-cone formulation that are relevant to Carrollian physics. The Carroll group, which arises as the ultrarelativistic (speed of light approaching zero) limit of the Poincar\'e algebra~\cite{Bacry:1968zf}, has been linked to symmetries of null hypersurfaces, the BMS symmetry of gravity, and flat-space holography (see~\cite{Duval:2014uva, Ciambelli:2019lap, Donnay:2022aba, Bergshoeff:2022eog} and references therein).
\par 
The light-cone formulation of field theories is based on Dirac's front form of relativistic dynamics~\cite{Dirac:1949cp}, where the time evolution of a system occurs along a lightlike or null direction. Light-cone physics rose to prominence following the seminal work of Weinberg~\cite{Weinberg:1966jm} in the late sixties. Weinberg showed that certain problematic Feynman diagrams in quantum field theories, such as vacuum fluctuations, take a simpler form in the light-cone or infinite-momentum frame, leading to finite or zero contributions. This simplicity was later attributed to an underlying three-dimensional Galliean invariance, which gives the four-dimensional theories a nonrelativistic structure~\cite{Susskind:1967rg}. On the other hand, the light-cone coordinate system comprises two null hypersurfaces that possess a Carrollian structure. While the Galilean or nonrelativistic nature of theories in the light-cone formulation has received considerable attention in the past~\cite{Susskind:1967rg, Kogut:1969xa, Toth:1974sb, Leutwyler:1977vy}, their Carrollian properties have not been discussed in much detail.  Hence, this article aims to connect some aspects of the light-cone formulation to Carrollian physics.

\par We emphasize the double-null nature of the light-cone coordinates, which corresponds to \textit{a choice of time} within this framework. The freedom to choose the time coordinate gives the light-cone Poincar\'e algebra a unique structure. As a result, the light-cone Poincar\'e generators obey various kinematical Lie algebras in one lower spacetime dimension. Of particular interest are the subalgebras of Galilei, Carroll, and Bargmann types, and their physical relevance in the light-cone formulation. Having discussed these subalgebras,  we will focus on the Carrollian features of field theories in the light-cone formalism. In recent years, Carrollian field theories have emerged as potential candidates for flat space holography, alongside celestial conformal field theories. In the standard approaches to Carrollian field theories, the ultrarelativistic limit of Lorentzian field theories leads to two classes of theories, namely the electric and the magnetic types. We show that field theories on the light front have a magnetic Carroll Hamiltonian. Thus, we propose a straightforward recipe for deriving magnetic Carroll actions from Lorentzian light-cone actions without using any group contraction method.

\section{Choice of light-cone time}
Starting with $(d+1)$-dimensional Minkowski spacetime in Cartesian coordinates, $x^\mu = (x^0, x^1 ,..., x^d) $, we consider two null vectors
\begin{equation}
m^\mu = \frac{1}{\sqrt 2}(-1 ,0 ,0, ... ,1)\,, \quad n^\mu =\frac{1}{\sqrt 2}(-1 ,0 ,0, ... ,-1) \,,
\end{equation}
such that ${n.n} = m.m =0$ and $n.m = -1$. The light fronts are the null planes obtained by projecting along the two null vectors
\begin{eqnarray} \label{LCcoord}
x^+ = m.x = \frac{1}{\sqrt 2}(x^0 + x^d) \,,&& x^- = n.x= \frac{1}{\sqrt 2}(x^0 - x^d) \,.
\end{eqnarray}
$m$ and $n$ are normals to the null planes $x^+=0$ and $x^-= 0$ respectively.  In the light-cone coordinates, the line element in Minkowski spacetime reads~\footnote{By `light-cone' or `light-front' coordinates, we always refer to planar null coordinates defined in~\eqref{LCcoord}, and not the retarded and advanced time coordinates, $t \pm r$.}
\begin{equation} \label{LCds2}
dS^2= -2 dx^+ dx^- + \delta_{ij}dx^i dx^j\,, \quad (i,j = 1,2, ..., d-1)\,.
\end{equation}
\par
\begin{figure}[ht]
	\begin{center}
	\includegraphics[width= 3.0in]{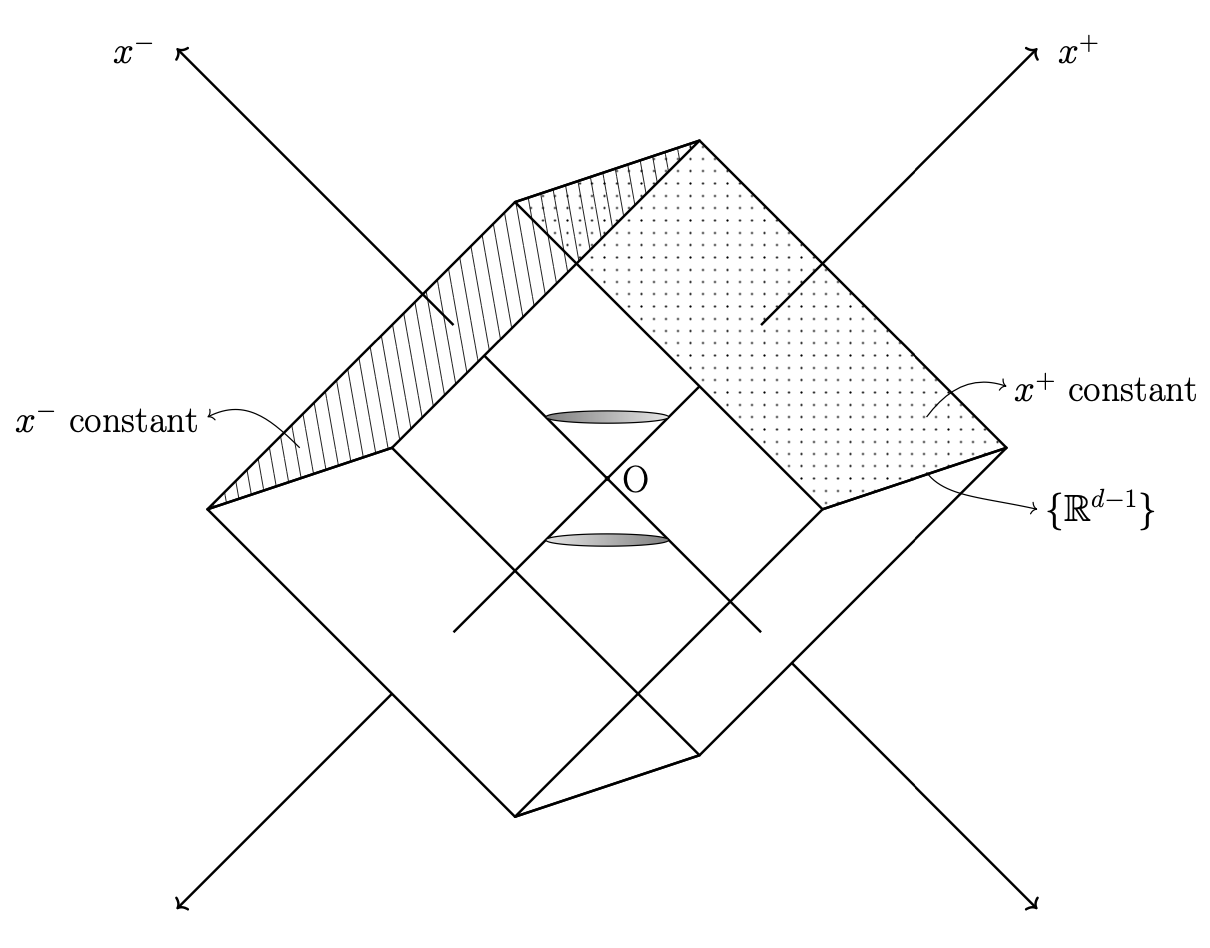}
\end{center}
		\caption{{The double-null light-cone coordinate system} } \label{fig:LCcoord}
\end{figure}

An interesting feature to note  from the line element is the double-null nature of the light-cone coordinates, which corresponds to a $\mathbb R^{d-1} \times \mathbb R \times \mathbb R$ split of the Minkowski spacetime, as shown in Fig~\ref{fig:LCcoord}. This hints at an apparent `duality' between the light-cone coordinates $x^\pm$ $-$ a choice of time $-$ since either of the two may be treated as the time coordinate. In most cases, such as low-energy effective descriptions and scattering amplitudes techniques~\cite{Brink:1982wv,Bengtsson:1983pd,Bengtsson:1983pg,Gorsky:2005sf,Mansfield:2005yd,Ananth:2007zy}, it suffices to pick one of the light-cone coordinates as time for the entirety of the analysis while treating the other one as a spatial coordinate. However, certain physical problems, particularly those pertaining to the initial value problem and quantization of massless fields on null planes, require both light-cone directions to be treated as time~\cite{Steinhardt:1979it,Mccartor:1988bc,Heinzl:1993px,Barnich:2024aln}. In these instances, the $x^\pm$ duality, \ie, the fact that both can play the role of the light-cone time, is not merely a choice but a necessity for a complete understanding of the quantization properties of a physical system. 
\par 
In this article, we will shed new light on the $x^\pm$ duality from an algebraic point of view. Following the works of~\cite{Duval:1984cj, Duval:2014uoa}, we shall discuss the concept of two non-Einsteinian times $-$ Galilean (or Newtonian) time and Carrollian time, adapted to the light-cone framework. Depending on which light-cone coordinate is taken to be the Newtonian time, one finds two distinct sets of Galilei, Carroll, and Bargmann subalgebras within the Poincar\'e algebra, which are mapped into one another upon exchanging $x^+$ with $x^-$.

\section{Aspects of light-cone Poincar\'e algebra}
We begin by reviewing some important features of the light-cone Poincar\'e algebra (see~\cite{Leutwyler:1977vy} for more details). The generators $(P_\mu, M_{\mu\nu})$ in light-cone coordinates satisfies the usual Poincar\'e algebra in $(d+1)$ dimensions
\bea
\left[P_\mu\,, P_\nu \right] &=& 0\,, \\
\left[M_{\mu \nu}\,, P_\rho \right] &=& \eta_{\nu\rho}P_{\mu} - \eta_{\mu \rho} P_{\nu}\,, \\
\left[ M_{\mu\nu}\,, M_{\rho \sigma}\right] &=& \eta_{\mu \sigma} M_{\nu \rho} + \eta_{\nu \rho} M_{\mu \sigma} -  \eta_{\mu \rho} M_{\nu \sigma} - \eta_{\nu \sigma} M_{\mu \rho} \,,
\eea
where the greek indices $\mu, \nu, ...$ now run over $(+, -, i)$ and $\eta_{\mu\nu}$ stands for the light-cone Minkowski metric.
For our convenience, we relabel the generators spanning the light-cone Poincar\'e algebra $\mathfrak p$ as follows
\begin{eqnarray}
P_+ = E\,,&  P_- = \eta\,, & ~~P_i \,, \nn\\
M_{+i} = K_i\,, & ~~M_{-i} = B_{i}\,,&~~ M_{ij}\,, \quad M_{+-}= D \,.
\end{eqnarray}
An element of the light-cone Poincar\'e algebra $\mathfrak p$ is given by
\begin{eqnarray}
X_{\mathfrak p}&=& a^+ E+ a^- \eta + a^i P_i +\beta D+ \gamma^i K_i + \sigma^i B_i + \frac{1}{2}\omega^{ij}M_{ij}  \nonumber \\
&=& (a^+ - \beta x^+ + \gamma_i x^i)\frac{\del}{\del x^+} +( a^- +\beta x^- +\sigma_i x^i) \frac{\del}{\del x^-} \nonumber \\
&&+\left(\omega^{i}{}_j x^j + \sigma^i x^+ +\gamma^i x^- + a^i\right)\frac{\del}{\del x^i}  \,, 
\end{eqnarray}
where we have relabeled the Lorentz parameters, $\omega^{+-},\omega^{+i}$ and $\omega^{-i}$, as $\beta, \gamma^i$ and $\sigma^i$, respectively. The generator $M_{+-}$, which we denote by $D$, acts as a dilatation operator that scales the $x^+$ and $x^-$ directions while keeping $x^i$ fixed. Under a dilatation $D$, a light-cone operator $\mathcal O$ transforms as 
\begin{equation}
	[D, \mathcal O] = \alpha \mathcal O\,,
\end{equation}
where the eigenvalue $\alpha$, often referred to as the ``goodness value'' ~\cite{Leutwyler:1977vy}, is an integer.
We can classify all the light-front Poincar\'e  generators according to their transformation under $D$, as shown in Table~\ref{table:LCgen}.

 \begin{table}[h]
   \begin{center}
 \caption{Generators of light-cone Poincar\'e algebra} \label{table:LCgen}
 	\begin{tabular}{c c c c c} 
 		\hline
 		$\alpha$ & $P_\mu$ & $M_{\mu\nu}$ & & Subgroups $G_{\alpha}$ \\ [0.5ex] 
 		\hline\hline
 		1 & $\eta$ & $B_i  $ & & $G_{1}$ \\ 
 		0 & $P_i$ & $M_{ij}\,,$&$D $& $G_0$ \\
 		-1 & $E$  &$ K_i$&  & $G_{-1}$ \\
 		\hline
 	\end{tabular}
	 \end{center}
	 \end{table}
Both $G_{+1}$ and $G_{-1}$ are abelian ideals of the light-cone Poincar\'e algebra. Additionally, there are two subgroups with a semi-direct product structure, 
\begin{eqnarray} \label{LCkin}
	\mathbb S_{+}&=& G_{0} \ltimes G_{1}  ~=~ \{ \eta, P_i, M_{ij}, B_i, D\} \nonumber \\
	\mathbb S_{-}&= &G_{0} \ltimes G_{-1} ~=~ \{ E, P_i, M_{ij}, K_i, D\}\,.
\end{eqnarray}
The subgroups $\mathbb S_{\pm}$ are the kinematical groups or stability groups of the null planes $x^{\pm}= 0$. While the subgroup $\mathbb S_{+}$ (or $\mathbb S_-$) gives the kinematics  within the null plane, the remaining generators comprising the abelian ideal $G_{-1}$ (or $G_{+1}$) form the dynamical part of the algebra that governs the evolution of the physical system as a function of null time, $x^+$ (or $x^-$).
\par For the sake of completeness, we list all the commutators~\footnote{Here, we use the convention that the first entry in a commutator $[X_{1}, X_2]$ is taken from the column and the second one from the row, \eg $[E, D] = E$ in the first row.} of the light-cone Poincar\'e algebra $\mathfrak{p}$ in Table~\ref{table:Poincare}.  

 \begin{table}[h]
 \begin{center}
 	\caption{Commutators of light-cone Poincar\'e algebra $\mathfrak p$}
\label{table:Poincare}
 	\footnotesize
	\begin{tabular}{c|c c c c c c c}
		\hline
		 & $E$ & $\eta$ &$P_i$ & $D$ & $M_{ij}$ & $B_i$ & $K_i$  \\[0.5ex] 
		 	\hline \hline
		 $E$ &$0$ &$0$ &$0$ &$E$ &$0$ &$-P_i$ & $0$ \\
		 $\eta$ &$0$& $0$ & $0$ & $-\eta$ & $0$ & $0$ & $-P_i$ \\
		 $P_l$ &$0$& $0$ & $0$ & $0$   &$\delta_{l[i}P_{j]} $ & $-\delta_{il}\eta$ & $-\delta_{il}E$ \\
		  $D$ & $-E$&$\eta$ &$0$ &$0$ &$0$ &$B_i$ &$-K_i$ \\
		   $M_{lm}$ &$0$&$0$ &$-\delta_{i[l}P_{m]}$ &$0$ & $\delta_{[i[m}M_{l]j]}$& $\delta_{i[m} B_{l]} $  & $\delta_{i[m} K_{l]} $ \\
		    $B_l$ &$P_l$ &$0$ &$\delta_{il} \eta$ &$-B_l$ & $-\delta_{l[i}B_{j]}$ & $0$&$M_{li}+\delta_{il}D$ \\
		    $K_l$&$O$ &$P_l$ &$\delta_{il}E$ &$K_l$ & $-\delta_{l[i}K_{j]}$ & $-M_{il}-\delta_{il}D$&$0$ 	\\[0.5ex] 
	   \hline
	\end{tabular}
	\normalsize
		\end{center}
	\end{table}

\par In four spacetime dimensions, the kinematical part of the light-cone Poincar\'e algebra is a seven-parameter subgroup as opposed to the usual six-parameter kinematical subgroups found in the instant form of dynamics. Hence, a key advantage of working in this formulation is that the light fronts possess the largest kinematical subgroup among all admissible choices of initial hypersurfaces for Hamiltonian theories. This feature tremendously simplifies the unknown dynamical part of the problem, proving to be very effective for deriving interacting Hamiltonian actions in many instances, such as higher-spin theories and supersymmetric theories~\cite{Bengtsson:1983pd, Bengtsson:1983pg}.
\section{Carroll, Bargmann and all that}
Minkowski spacetime in light-cone coordinates exhibits a flat Bargmann structure~\cite{Duval:1984cj, Duval:2014uoa}, encompassing both Galilean and Carrollian spacetimes in one lower dimension. The Galilean spacetime follows from a null dimensional reduction \`a la Kaluza-Klein, while the Carrollian spacetime may be viewed as an embedded null hyperplane~\cite{Gauntlett:1990xq,Julia:1994bs,Duval:2014uoa}. The Bargmannian structure, combined with the choice of light-cone time, gives rise to a rich array of interesting kinematical Lie algebras within its Poincar\'e algebra.

\subsection{Kinematical Lie algebras}
Before exploring the various subalgebras of the light-cone Poincar\'e algebra, we briefly review a few basic elements of kinematical Lie algebras (see~\cite{Figueroa-OFarrill:2017sfs} for more details). The key ingredients of a $(n+1)$-dimensional kinematical Lie algebra are the generators
\begin{equation}
	\mathcal{K} = \{L_{ab}, B_a, P_a, H\} \,, \quad a,b,c,.. = 1,2,... , n
\end{equation}
where the generators
\begin{itemize}
	\item $L_{ab}$ span the $\boldsymbol{so}(n)$ algebra
	\begin{equation} \label{kin1}
		[\boldsymbol{L}, \boldsymbol{L}]= \boldsymbol{L}\,,
	\end{equation}
	\item $B_a$ and $P_a$ transform as vectors under $L_{ab}$
	\be \label{kin2}
	[\boldsymbol{L}, \boldsymbol{B}]= \boldsymbol{B} \,, \quad 		[\boldsymbol{L}, \boldsymbol{P}]=\boldsymbol{P} \,,
\ee
	\item and, $H$ is a scalar
	\begin{equation} \label{kin3}
			[\boldsymbol{L}, H]= 0\,.
	\end{equation}
\end{itemize}
{In the commutation relations above, we denote the $\mathfrak{so}(n)$ generators and the vectors with bold letters, while suppressing the $\mathfrak{so}(n)$ indices on them.}
The vectors $\boldsymbol{B}$ and $\boldsymbol{P}$, here, should not be confused with the light-cone Poincar\'e generators  $B_i$ and $P_i$ discussed in the previous section. These $\mathfrak{so}(n)$ vectors may be identified with the Poincar\'e generators in some instances, but that is not always the case.

\par The kinematical Lie algebras are, then, defined through the non-zero commutators among $H,\boldsymbol{B},$ and $ \boldsymbol{P}$. In particular, the Galilei, Carroll and Bargmann algebras are defined as follows
\bea
\text{Galilei}\ \mathfrak g : && [H, \boldsymbol{B}] =- \boldsymbol{P} \label{Gall}\,, \\
\text{Carroll}\ \mathfrak c : && [\boldsymbol{B}, \boldsymbol{P}] = H \label{Carr} \,,  \\
\text{Bargmann}\ \mathfrak b : &&  [H, \boldsymbol{B}] =- \boldsymbol{P}\,, \quad [\boldsymbol{B}, \boldsymbol{P}] = Z \label{Barg} \,, 
\eea
where $Z$ is a central element. Thus, the Carroll and Galilei groups may be unified into a larger Bargmann group~\cite{Duval:2014uoa, Figueroa-OFarrill:2022pus}.
 \subsection{Subalgebras of light-cone Poincar\'e algebra}
 
 \par 
Due to a $\mathbb R^{d-1} \times \mathbb R \times \mathbb R$ split of the Minkowski spacetime, instead of the usual $\mathbb R^{d} \times \mathbb R$, the light-cone Poincar\'e generators split as
\begin{itemize}
	\item $M_{ij}$ spanning an $\mathfrak{so}(d-1)$ algebra, 
		\item Three vectors $\{P_i$, $B_i$, $K_i\}$ under the $\mathfrak{so}(d-1)$, and 
			\item Three scalars $\{E, \eta, D\}$ under the $\mathfrak{so}(d-1)$.
			\end{itemize}
Among these generators, one can find two subsets that satisfy the conditions \eqref{kin1}, \eqref{kin2} and \eqref{kin3}, along with~\eqref{Barg}, and hence, form Bargmann subalgebras~\footnote{We shall use the subscript `$+$' or `$-$' to label the algebras in order to specify which light-cone coordinate, $x^+$ or $x^-$, is chosen as the Newtonian time.} 
\bea
\mathfrak b_+ &=& \{E, \eta, M_{ij}, B_i, P_i\} \,, \quad \eta \ \text{central element} \,,\\
\mathfrak b_- &=& \{E, \eta, M_{ij}, K_i, P_i\} \,, \quad E \ \text{central element} \,.
\eea
By assigning a notion of Newtonian and Carrollian time~\cite{Duval:2014uoa} to the light-cone coordinates $x^\pm$ within each Bargmann group, we can identify a Galilei and a Carroll subgroup. The important point here is that  we have at our disposal, two vectors, $B_i$ and $K_i$, that can play the role of the Galilean (or Carrollian) boosts, and two scalars, $\eta$ and $E$, that centralize the corresponding Galilean algebra. We elaborate on this point below.
\par
We can choose one of the null coordinates to be the `Newtonian'  time for the Galilei group. The Hamiltonian (energy) is then given by the momentum corresponding the Newtonian time. For instance, if $x^+$ is taken to be the Newtonian time, the generator $E= P_+$ becomes the Hamiltonian. As a result, the momentum associated with the other null coordinate $x^-$ appears as a nonrelativistic mass, $\eta= P_-$ in the mass-shell condition 
\be
2E \eta - P^i P_i = 0 \quad \Rightarrow  \quad H_{\mathfrak g}= E= \frac{P^2}{2\eta}\,.
\ee 
This null coordinate may then be treated as a `Carrollian' time. In the Bargmann algebra $\mathfrak{b}_+$, the generator $\eta$ indeed appears as a central element. Thus, treating $x^+$ as Newtonian time and $x^-$ as Carrollian, we can readily identify the Galilei and Carroll group contained in $\mathfrak{b}_+$ as
\bea
&& \text{Galilei:} \quad \mathfrak{g}_+ = \{E, M_{ij}, B_{i}, P_{i}\} \,, \\
&& \text{Carroll:} \quad \mathfrak{c}_+ = \{\eta, M_{ij}, B_i, P_i\} \,.
\eea
The 3-dimensional Galilei subalgebra within the 4-dimensional Poincar\'e algebra discussed in the light-cone literature~\cite{Susskind:1967rg, Kogut:1969xa, Toth:1974sb, Gomis:1978mv},  indeed coincides with $\mathfrak{g}_+$. It was also pointed out that $P_
-$ centralizes this algebra, but the resulting centrally-extended algebra was not identified as a Bargmann algebra $\mathfrak b_+$.
\par
Now, if we instead take $x^-$ to be the Newtonian time and $x^+$ to be Carrollian, we find another set of Galilei and Carroll groups contained in the other Bargmann group $\mathfrak{b}_-$ as follows
 \bea
 && \text{Galilei:} \quad \mathfrak{g}_- = \{\eta, M_{ij}, K_{i}, P_{i}\} \,, \\
 && \text{Carroll:} \quad \mathfrak{c}_- = \{E, M_{ij}, K_i, P_i\} \,.
 \eea
 \par The Carroll subgroups $\mathfrak{c}_{\pm}$ are the stability groups of the $d$-dimensional light fronts, $x^\pm = \text{constant}$, embedded in a $(d+1)$-dimensional Minkowski spacetime (see Fig~\ref{fig:LCcoord}). Note that in four dimensions, the stability groups $\mathbb S_{\pm}$, given in~\eqref{LCkin}, are seven-dimensional only for the light fronts at $x^\pm =0$, where the generator $M_{+-}$ is kinematical. In all other cases, the stability groups of the null fronts are given by $\mathfrak{c}_{\pm}$, which are six-dimensional.
 \par
{Interestingly, the Lorentz generators in $\mathfrak p$ satisfy another kinematical Lie algebra characterized by the relations
\be
[H, \boldsymbol{B}] = \boldsymbol{B}\,,\quad  [H, \boldsymbol{P}] = -\boldsymbol{P}, \quad  [\boldsymbol{B}, \boldsymbol{P}] = H+\boldsymbol{L}\,,
\ee
with the following identification: $
H = D\,, \, \boldsymbol{L} = M_{ij} \,, \, \boldsymbol{B}= B_i\,, \,  \boldsymbol{P} = K_i \,. $
This Lie algebra is geometrically associated with a special class of homogenous spaces, named the `Carrollian light cone' in~\cite{Figueroa-OFarrill:2018ilb}.}
\par
Therefore, in $(d+1)$ spacetime dimensions, the light-cone Poincar\'e algebra $\mathfrak p$, spanned by $ dim(\mathfrak p) =\frac{(d+1)(d+2)}{2}$ generators, contains the $d$-dimensional subalgebras $(\mathfrak b_{\pm}, \mathfrak c_\pm, \mathfrak g_{\pm})$ with a group dimension
\bea
&&dim(\mathfrak{b}_{\pm})= \frac{d(d+1)}{2}+1 \,, \\
	&& dim(\mathfrak{g}_{\pm}) =  dim(\mathfrak{c}_{\pm})=  \frac{d(d+1)}{2} \,.
\eea

\par
We summarize these subalgebras in Table~\ref{table:subgroups}, highlighting the central elements and key commutators that define them, while  all other commutation relations can be found in Table~\ref{table:Poincare}.  
\begin{table}[h]
 \begin{center}
\caption{Kinematical Lie subalgebras within light-cone Poincar\'e algebra}
\label{table:subgroups}
 	\footnotesize
	\begin{tabular}{c |c| c } 
		\hline
		Time & $x^+$ Newtonian, $x^-$ Carrollian &$x^+$ Carrollian, $x^-$ Newtonian \\ [0.5ex] 
			\hline\hline
		Bargmann $\mathfrak b$ &$\mathfrak b_+ = \{E, \eta, M_{ij}, B_i, P_i\}$ &  $\mathfrak b_- = \{\eta, E, M_{ij}, K_i, P_i\}$ \\
		& $\eta$ central element & $E$ central element \\
		\hline
		 Galilei $\mathfrak g$ &$\mathfrak g_+ = \{E, M_{ij}, B_i, P_i\} $  & $\mathfrak g_- = \{\eta, M_{ij}, K_i, P_i\} $\\
		 & $[E, B_i]= -P_i$& $[\eta, K_i]= -P_i$\\
		 & Hamiltonian $H_{\mathfrak g_+} = E$& Hamiltonian $H_{\mathfrak g_-} = \eta$\\
		 \hline 
		 Carroll $\mathfrak c $& $\mathfrak c_+ =\{\eta, M_{ij}, B_i, P_i\}$ &   $\mathfrak c_- =\{E, M_{ij}, K_i, P_i\}$  \\
		 &  $[B_i, P_j]= \delta_{ij}\eta$ & $[K_i, P_j]= \delta_{ij}E$ \\
		 & Hamiltonian $H_{\mathfrak c_+} = \eta$ &  Hamiltonian $H_{\mathfrak c_-} = E$ \\
		 \hline 
		\end{tabular}
\normalsize
\end{center}
\end{table}

 As alluded to before, the two null coordinates, $x^+$ and $x^-$, are completely interchangeable in the light-cone coordinate system. This freedom to choose the light-cone time is also reflected in these subgroups. On swapping $x^+ $ with $x^-$,  the roles of certain generators in $\mathfrak p$ are exchanged
\bea
x^+ \longleftrightarrow x^-\,, \quad E \longleftrightarrow \eta\,, \quad K_{i} \longleftrightarrow B_i \,,
\eea
which results in the two sets of Bargmann, Galilei and Carroll subgroups being mapped into each other
\be
\mathfrak b_+ \longleftrightarrow \mathfrak b_-\,, \quad \mathfrak g_+ \longleftrightarrow \mathfrak g_- \,,\quad\mathfrak c_+ \longleftrightarrow\mathfrak c_{-} \,.
\ee
%

\par
\textit{The key point is that these subalgebras are inherent to the light-cone Poincar\'e algebra and do not arise from a nonrelativistic or ultrarelativistic limit.} In the light-cone framework, these non-Lorentzian symmetries $-$ Carrollian and Galilean $-$ can coexist in the same physical system, albeit as subalgebras of the Poincar\'e symmetry in one higher dimension.

\section{Magnetic Carroll actions on the light front}
In this section, we explore the Carrollian nature of field theories on the light front. In particular, we illustrate that the Hamiltonian density derived from light-cone actions are of the magnetic Carroll type. For simplicity, we restrict our discussion to the case of scalar field theory in $(d+1)$ dimensions. 
\par 
The massless scalar field action
\be
\mathcal S[\phi, \dot{\phi}] = \int d^{d+1} x \, \mathcal L\,, \quad \mathcal L = -\frac{1}{2} \eta^{\mu \nu} \del_\mu \phi \del_\nu \phi \,,
\ee 
in the light-cone coordinates, reads
\bea \label{LC-Lag}
\mathcal S_{L} = \int dx^+ dx^- d^{d-1}x \, \left( \del_+ \phi \del_- \phi - \frac{1}{2} \del_i \phi \del^i \phi
 \right) \,.
\eea
We choose $x^+$ as the Carrollian time for the light-front Hamiltonian analysis. This choice aligns with existing light-cone results, where $x^+$ is conventionally treated as time. A notable example is the BMS algebra in light-cone gravity, which can be obtained as a local extension of the Poincar\'e algebra~\cite{Ananth:2020ngt} as well as a conformal extension of the Carroll algebra $\mathfrak c_-$~\cite{Ananth:2020ojp}. One could instead consider a theory with $x^-$ as the Carrollian time, in which case the relevant symmetry group would be $\mathfrak c_+$, as described in Table~\ref{table:subgroups}.
\par
Treating $x^+$ as Carrollian time,  we define the conjugate momenta as
\be
\pi_\phi = \frac{\delta \mathcal L^{lc}}{\delta(\del_+ \phi)} = \del_- \phi \,.
\ee
Since the conjugate momenta involve no time derivates, we get a primary constraint from the definition of $\pi_\phi$
\be
\chi_\phi= \pi_\phi - \del_- \phi \,.
\ee
At this point, we choose to strongly impose the second-class constraint, effectively eliminating the conjugate momenta from the action, as is customary in the $lc_2$ formalism. The canonical Hamiltonian density so obtained involves the spatial derivatives $\del_i$ only
\be \label{LC-Ham}
\mathcal H^{lc} = \pi_\phi \del_+ \phi - \mathcal L^{lc} = \frac{1}{2} \del_i \phi \del^i \phi \,.
\ee
The light-cone Hamiltonian density exactly matches with the Hamiltonian density obtained in~\cite{Henneaux:2021yzg} for $d$-dimensional scalar field theory in the magnetic Carroll case. In fact, the Hamiltonian action, 
\be  \label{LCaction}
\mathcal S_H = \int dx^+ dx^- d^{d-1}x (\del_+ \phi \del_- \phi - \mathcal H^{lc})\,,
\ee
is the same as the Lagrangian action shown above, since the dynamics is now defined on the reduced phase space endowed with the Poisson (or more precisely, Dirac) bracket 
\be
\{ \phi(x), \phi(y) \} = \frac{1}{2}\, \varepsilon(x^--y^-)\, \delta^{(d-1)} (x-y) \,,
\ee
where $\varepsilon(x^--y^-)$ is the Heaviside step function.
\par \textit{The crucial point is that upon solving the second-class primary constraint to eliminate the conjugate momenta, one immediately obtains from the light-cone action a magnetic Carroll Hamiltonian in one lower dimension without taking any ultrarelativistic limit.} 
\par Alternatively, we could keep the conjugate momenta in the phase space and impose the constraint by means of a Lagrange multiplier. We will return to this point in Section~\ref{limits}. 

\subsection{Invariance under Carroll transformations}
We now focus on the Carroll transformations restricted to a constant $x^- $ surface, where the symmetry group of our interest is $\mathfrak c_-= \{ M_{ij}, P_i, K_i, E \} $
\bea \label{CarrTransf}
x^-& \rightarrow& x^- \nonumber \,, \\
x^+ & \rightarrow & x^+ + \gamma^i x_i + a^+ \nonumber \,, \\
x^i & \rightarrow & x^i + \omega^i_j x^j + \gamma^i x^- + a^i \,.
\eea
One can easily verify that the corresponding field transformations 
\be
\delta_{\mathfrak c_-} \phi = (\gamma^i x_i + a^+) \del_+ \phi + (\omega^i_j x^j + \gamma^i x^- + a^i) \del_i \phi \,.
\ee
render the light-cone action \eqref{LCaction} invariant. In the light-cone phase space, the canonical generators for the Carroll transformations are given in terms of the energy and momentum densities
\be
\mathcal{H}(x) = \frac{1}{2} \del_i \phi \,\del^i \phi\,,  \quad \mathcal P_i (x) = \del_- \phi\, \del_i \phi\,.
\ee
The commutation relations among the energy and momentum densities are 
\bea
\left[\mathcal H(x),\, \mathcal H(y) \right] &=& 0  \label{HH} \,, \\
\left[\mathcal H(x),\, \mathcal P_i (y)\right] &= &  \mathcal H(y) \, \delta(x^- - y^-) \, \del_i \delta^{d-1} (x-y) \,, \\
\left[ \mathcal P_i (x),\, \mathcal P_j(y) \right] &=& \frac{1}{2}  \big[ \del_j \delta^{d-1} (x-y) \mathcal P_i (y) + \del_i \delta^{d-1} (x-y) \mathcal P_j (x) \big].
\eea
From~\eqref{HH}, we note that the commutator of two Hamiltonian densities vanishes $-$ a signature of Carrollian dynamics~\cite{Teitelboim:1972vw,Henneaux:1979vn}. The canonical generators are then defined as
\bea \label{Carr-Gen}
&&E = \int dx^- d^{d-1}x \, \mathcal H(x) \,, \quad P_i = \int dx^- d^{d-1}x\, \mathcal P_i (x) \,, \\
&& M_{ij} = \int dx^- d^{d-1} x\, (x_i \mathcal P_j - x_j \mathcal P_i) \,,\\
&& K_i = \int dx^- d^{d-1} x \,(x_+ \mathcal P_i - x_i \mathcal H)\,.
\eea
The non-vanishing commutators of the Carroll algebra are
\bea \label{Carr-alg}
\left[ K_i ,\, P_j\right] &=& \delta_{ij} E \,, \\
\left[ M_{ij} ,\, P_l \right] &=& \delta_{lj} P_i - \delta_{il} P_j \,, \\
\left[ M_{ij} ,\, K_l \right] &=& \delta_{lj} K_i - \delta_{il} K_j \,, \\
\left[ M_{ij}, M_{lm} \right] &=& \delta_{im} M_{jl} - \delta_{il} M_{jm} - \delta_{jm} M_{il} + \delta_{jl}M_{im} \,.
\eea
\par Therefore, we see that the magnetic Carroll action for scalar field theory arises naturally in the light-cone formalism, when we solve the second-class constraints after reducing the theory to one lower dimension. However, this method does not seem to yield an electric Carroll action from the light-cone Lagrangian. To this end, we present below an alternative method for obtaining Carrollian actions on the light front, which closely resemble the conventional approaches to Carrollian field theories.
\subsection{Carroll limits of light-cone actions} \label{limits}
In the Hamiltonian formulation, the standard way to deal with constraints is through the introduction of Lagrange multipliers in the action~\cite{Bergmann:1949zz,Dirac:1950pj}. The constraints are then implemented in the extended phase space via the equations of motion for the Lagrange multipliers. We apply this procedure to the primary constraint obtained from the light-cone Lagrangian~\eqref{LC-Lag}. We augment the canonical Hamiltonian~\eqref{LC-Ham} with a term involving a Lagrange multiplier $\lambda$, resulting in the total Hamiltonian.
\be
\mathcal H_{T} = \mathcal H^{lc} + \lambda \chi_\phi = \frac{1}{2} \del_i \phi \del^i \phi + \lambda (\pi_{\phi} - \del_-\phi) \,.
\ee
In this extended phase space, the Hamiltonian action takes the form
\be
\mathcal S_{H} [\phi, \pi_{\phi}, \lambda ] = \int dx^- dx^+  d^{d-1}x  \Big( \pi_{\phi} \del_+ \phi - \frac{1}{2} \del_i \phi \del^i \phi - \lambda (\pi_{\phi} - \del_- \phi) \Big) \,.
\ee
The equation of motions obtained from the action are as follows
\bea \label{Sextended}
&\lambda : & \pi_{\phi} -\del_- \phi = 0\,, \\
& \pi_\phi :& \del_+ \phi - \lambda = 0 \,, \\
& \phi :& -\del_+ \pi_{\phi} + \del_- \lambda + \frac{1}{2} \del_i \del^i \phi = 0 \,.
\eea
The constraint $\chi_\phi$ is retrieved from the equation of motion for $\lambda$. The Hamiltonian densities at two points do not commute in this case, \ie\, $[\mathcal H_T(x),\, \mathcal H_T(y) ] \neq 0 $, indicating that the dynamics is not Carrollian.
\par We now restrict the theory to a constant $x^-$ surface by restricting the domain of $x^-$ to a finite width $\rho$, where $\rho$ is a small constant, using a delta function distribution~\cite{Chen:2023pqf}
\bea
\delta_\rho (x^- -x^-_0) &=& 
\begin{dcases}
      \frac{1}{\rho} \,, &x^-_0 - \frac{\rho}{2} < x^- < x^-_0 + \frac{\rho}{2}  \\
0 \,,& \text{otherwise} \,.
    \end{dcases}
\eea
In the limit $\rho \rightarrow 0$, we can devise two distinct rescalings of the canonical variables $\phi, \pi_{\phi}$ and $\lambda$, that lead to Carroll actions of the electric and magnetic types
 \bea
{}^{(d)}\mathcal S_{Carr} &\equiv& \lim_{\rho \to 0}  {}^{(d+1)}\mathcal S_{\rho} \nn \\
 & =& \lim_{\rho \to 0}   \int dx^+   d^\perp xdx^- \delta_\rho (x^- -x^-_0)  {}^{(d+1)} \mathcal L_H [\phi, \pi_{\phi}, \lambda ] \,.
 \eea
\vskip 0.2cm
\ndt 
\textit{Case I: Magnetic Carroll limit} 
\vskip 0.2cm
\ndt
We rescale the fields in the Hamiltonian action~\eqref{Sextended}  as follows
 \be
\phi \rightarrow \phi \,,\quad  \del_-\phi \rightarrow p_\phi\,, \quad \pi_{\phi} \rightarrow {\pi_\phi} \,, \quad \lambda \rightarrow \lambda \,.
\ee
We strongly impose the second-class constraint by setting $\pi_\phi = p_\phi$. Thus, as $\rho \rightarrow 0$, we recover the magnetic Carroll action~\eqref{LC-Ham}
\be
\mathcal S_H =\int dx^+ d^{d-1} x \left( p_\phi \del_+ \phi -\mathcal H^{mag}\right)\,,
\ee
with the \textit{magnetic Carroll} Hamiltonian and momentum densities
\be 
\mathcal H^{mag} = \frac{1}{2} \del_i \phi\del^i \phi \,, \quad \mathcal P_i^{mag} = \del_- \phi \del_i \phi\,.
\ee
The nontrivial Poisson bracket in this case is $\{\phi (x), p_{\phi} (y)\} = \delta^{d-1} (x-y)$. The canonical generators for the Carroll transformations are obtained from the ones defined in~\eqref{Carr-Gen}, which obey the Carroll algebra~\eqref{Carr-alg}, by projecting them onto the lower-dimensional null hypersurface. The Carroll transformation laws for the fields may be obtained as
\be
\delta_G \phi = \{ \phi\,, G_{\mathfrak c} \} \,, \quad \delta_G p_\phi = \{ p_\phi\,, G_{\mathfrak c} \} \,.
\ee
The invariance of the action under the Carroll transformations follows accordingly.
\par
One can verify that the magnetic Carroll theory so obtained is identical to the ones derived in~\cite{Henneaux:2021yzg}.
\vskip 0.2cm
\ndt
\textit{Case II: Electric Carroll limit}
\vskip 0.2cm
\ndt
We now consider a different rescaling of the fields with respect to $\rho$
\be
\phi \rightarrow \rho \phi \,, \quad \pi_\phi \rightarrow \frac{\pi_\phi}{\rho}\,, \quad \lambda \rightarrow \rho \lambda \,.
\ee
As $\rho \rightarrow 0$, the light-cone Hamiltonian action~\eqref{Sextended} becomes
\be
\mathcal S_{H} = \int dx^+ d^{d-1}x \left( \pi_\phi \del_+ \phi - \mathcal H^{elec} \right) \,,
\ee
with the \textit{electric Carroll} Hamiltonian and momentum densities
\be 
\mathcal H^{elec} =\lambda \pi_\phi \,, \quad \mathcal P_i^{elec} = \pi_{\phi} \del_i \phi \,.
\ee
However, this theory possesses an extra gauge symmetry generated by the first-class constraint, $C= \pi$
\be
G[\sigma] = \int dx^\perp \sigma \pi\,.
\ee
The transformation laws for the fields read
\be
\quad \delta_\sigma \phi = \sigma (x^+, x^i)\,, \quad \delta_\sigma \lambda = \del_+ \sigma (x^+,x^i)\,, \quad \delta_\sigma \pi =0 \,,
\ee
indicating that one can gauge away the `electric' Carroll field $\phi$ completely using this gauge freedom. Therefore, this pure-gauge theory does not describe electric Carroll scalars. To derive the electric Carroll theory, we must consider more general Hamiltonian dynamics in one higher dimension.
\vskip 0.3cm
\centerline{* \quad * \quad *}
\vskip 0.2cm
\par
We have discussed two methods for deriving magnetic Carroll actions from Lorentzian field theories, which essentially differ in how we impose the second-class constraints in the Hamiltonian formulation. In the first case, we solve the constraint to eliminate the conjugate momenta, resulting in a Hamiltonian with only spatial derivatives. In the second case, the constraint is implemented through a Lagrange multiplier, and the Carrollian actions are obtained by suitably rescaling the canonical variables upon reduction to a null hypersurface.
\par Evidently, the magnetic Carroll Hamiltonian appears more naturally in the light-cone formalism than its electric counterpart. This contrasts with the standard approaches to Carrollian field theories, where the electric Carroll action follows from a simple rescaling of the fields, while the magnetic limit is more subtle. The crucial difference lies in the fact that light-cone Lagrangians are first-order in time derivatives. Hence, one always obtains some primary constraints from the definition of the conjugate momenta. Solving these constraints eliminates the kinetic term in the Legendre transform, $\mathcal{H} = \pi \dot{\phi} - \mathcal{L}$, leaving a canonical Hamiltonian with only spatial derivatives, hence rendering it magnetic Carroll.
 \par 
In a nutshell, if our goal is to find a magnetic Carroll action for a theory in $d$ dimensions, a straightforward approach would be to begin with the Lorentzian action in one higher dimension $(d+1)$ in the light-cone formulation and employ the first method. By treating one of the null coordinates as the Carrollian time, we can solve the second-class constraints obtained from the Lagrangian and work with Dirac brackets in the reduced phase space, thereby arriving at a magnetic Carroll Hamiltonian action.
\par
In order to obtain Carrollian actions for gauge theories, such as light-cone electromagnetism or gravity, we need an additional ingredient: a choice of gauge that effectively restricts the vector (or tensor) fields to one of the light fronts. We wish to address this problem in future work. 
\section*{Acknowledgements}
 We are grateful to Glenn Barnich and Simone Speziale for the initial discussions that led to this project. We thank Oscar Fuentealba and Marc Henneaux for fruitful discussions during the thematic program ``Carrollian Physics and Holography'' at the Erwin Schr\"odinger Institute, where a part of this work was completed. We acknowledge the support of the grant ID\# 62312 from the John Templeton Foundation, as part of the project {``The Quantum Information Structure of Spacetime'' (QISS)}.

\end{document}